\documentclass[fleqn,usenatbib]{mnras}
\usepackage{graphicx}	
\usepackage{amsmath}	
\usepackage{subcaption}
\usepackage{multirow}
\usepackage{amsmath}
\usepackage{url}
\usepackage{setspace}
\usepackage{longtable}
\usepackage{pdflscape}
\usepackage{float}
\usepackage{booktabs}
\usepackage{tabularx}
\usepackage{amssymb}

\usepackage{siunitx} 
\usepackage{wasysym}
\usepackage{rotating}
\usepackage{adjustbox}
\usepackage{xspace}
\usepackage{xcolor}
\usepackage[T1]{fontenc}
\usepackage{adjustbox}
\usepackage{enumitem,calc}
\usepackage{newtxtext,newtxmath} 

\newcommand{\kms}{kms$^{-1}\,$}

\newcommand{\hi}{{H\,{\small I}}\xspace }
\newcommand{\HI}{$\textbf{H}\,\scriptstyle\mathbf{I}$}
\newcommand{\mhi}{$M_\mathrm{HI}$} 
\newcommand{\dhi}{$D_\mathrm{HI}$} 

\newcommand{\pcm}{\,cm$^{-2}$}	
\newcommand\solMassPcSquare{$\rm M_{\odot}\,pc^{-2}$}

\title[The \dhi\ -- \mhi\ relation over the last gigayear]{MIGHTEE-\HI{}: the \HI{} Size--Mass relation over the last billion years}

\author[S. H. A. Rajohnson et al.]{Sambatriniaina H. A. Rajohnson$^{1}$,\thanks{E-mail: aychasam@gmail.com,}
	Bradley S. Frank$^{2,3,1}$,
	Anastasia A. Ponomareva$^{4}$, 
	Natasha Maddox$^{5}$,
	\newauthor{Renée C. Kraan-Korteweg$^{1}$,
		Matt J. Jarvis$^{4,6}$,
		Elizabeth A. K. Adams$^{7,8}$,
		Tom Oosterloo$^{7,8}$, 
		Maarten Baes$^{9}$,}
	\newauthor{Kristine Spekkens$^{10}$,
		Nathan J. Adams$^{11}$, 
		Marcin Glowacki$^{12,13}$,
		Sushma Kurapati$^{1}$,
		Isabella Prandoni$^{14}$}
	\newauthor{Ian Heywood$^{4,15,16}$,
		Jordan D. Collier$^{3,17}$, 
		Srikrishna Sekhar$^{3,18,19}$
		and Russ Taylor$^{3,19}$}\\
	$^{1}$Department of Astronomy, University of Cape Town, Private Bag X3, Rondebosch 7701, South Africa\\
	$^{2}$South African Radio Astronomy Observatory, 2 Fir Street, Observatory, 7925, South Africa\\
	$^{3}$The Inter-University Institute for Data Intensive Astronomy (IDIA), and University of Cape Town, Private Bag X3, Rondebosch, 7701, South Africa\\
	$^{4}$Astrophysics, Department of Physics, University of Oxford, Keble Road, Oxford OX1 3RH, UK\\
	$^{5}$Faculty of Physics, Ludwig-Maximilians-Universit\"at, Scheinerstr. 1, 81679 Munich, Germany\\
	$^{6}$Department of Physics and Astronomy, University of the Western Cape, Robert Sobukwe Road, 7535 Bellville, Cape Town, South Africa\\
	$^{7}$ASTRON, the Netherlands Institute for Radio Astronomy, Oude Hoogeveesedijk 4, 7991 PD Dwingeloo, The Netherlands\\
	$^{8}$Kapteyn Astronomical Institute, PO Box 800, 9700 AV Groningen, The Netherlands\\
	$^{9}$Sterrenkundig Observatorium Universiteit Gent, Krijgslaan 281 S9, B-9000 Gent, Belgium\\
	$^{10}$Department of Physics and Space Science, Royal Military College of Canada, PO Box 17000, Station Forces, Kingston, K7K 7B4, Canada\\
	$^{11}$Jodrell Bank Centre for Astrophysics, University of Manchester, Oxford Road, Manchester, UK\\
	$^{12}$International Centre for Radio Astronomy Research, Curtin University, Bentley, WA 6102, Australia\\
	$^{13}$Inter-University Institute for Data Intensive Astronomy, Bellville 7535, South Africa\\
	$^{14}$INAF-IRA, Via P. Gobetti 101, 40129, Italy\\
	$^{15}$Centre for Radio Astronomy Techniques and Technologies, Department of Physics and Electronics, Rhodes University, PO Box 94, Makhanda, 6140, South Africa\\
	$^{16}$South African Radio Astronomy Observatory, 2 Fir Street, Black River Park, Observatory, Cape Town, 7925, South Africa\\
	$^{17}$School of Science, Western Sydney University, Locked Bag 1797, Penrith, NSW 2751, Australia\\
	$^{18}$National Radio Astronomy Observatory, 1003 Lopezville Road, Socorro, NM 87801, USA\\
	$^{19}$Inter-University Institute for Data Intensive Astronomy, and Department of Astronomy, University of the Western Cape, Bellville, South Africa\\
}

\date{Accepted 2022 March 9. Received 2022 March 8; in original form 2021 December 7}

\pubyear{2022}

\begin{document}
\label{firstpage}
\pagerange{\pageref{firstpage}--\pageref{lastpage}}
\maketitle

\begin{abstract}
We present the observed \hi{} size--mass relation of 204 galaxies from the MIGHTEE Survey Early Science data. The high sensitivity of MeerKAT allows us to detect galaxies spanning more than 4 orders of magnitude in \hi{} mass, ranging from dwarf galaxies to massive spirals, and including all morphological types. This is the first time the relation has been explored on a blind homogeneous data set which extends over a previously unexplored redshift range of $0 < z < 0.084$, i.e. a period of around one billion years in cosmic time. The sample follows the same tight logarithmic relation derived from previous work, between the diameter (\dhi) and the mass (\mhi) of \hi{} discs. We measure a slope of 0.501$\pm$ 0.008, an intercept of $-3.252^{+0.073}_{-0.074}$, and an observed scatter of 0.057 dex. For the first time, we quantify the intrinsic scatter of $0.054 \pm 0.003$ dex (${\sim} 10 \%$), which provides a constraint for cosmological simulations of galaxy formation and evolution. We derive the relation as a function of galaxy type and find that their intrinsic scatters and slopes are consistent within the errors. We also calculate the \dhi\,--\,\mhi ~relation for two redshift bins and do not find any evidence for evolution with redshift. 
These results suggest that over a period of one billion years in lookback time, galaxy discs have not undergone significant evolution in their gas distribution and mean surface mass density, indicating a lack of dependence on both morphological type and redshift.

\end{abstract}


\section{Introduction}
\label{sec:intro}
Galaxies are gravitationally bound systems of stars, gas and dark matter. The processes governing their full gas cycle remain an area of active research. Theories of galaxy formation and evolution predict the gas infall onto galaxies to be the main mechanism supporting star formation and galaxy growth \citep{Giovanelli1988}. Galaxies must continuously accrete gas from an external environment to maintain their observed levels of star formation \citep[e.g.][]{Keres2005,Sancisi2008,Kauffmann2010}. Consequently, the environment in which a galaxy resides affects its evolution and thus its stellar mass and morphology \citep[see][]{Baldry2006,Peng2010}. Dense environments, for example, not only prevent further accretion of gas and ongoing star formation, but also play the crucial role in active gas stripping and subsequent loss of gas from galaxies. This can be observed through the morphology-density relation described in \cite{Dressler1997}, where high-density environments are populated by early-type galaxies (E, S0) whereas low-density environments are more dominated by late-type galaxies (S, Irr).

In addition to the morphology-density relation, other relations that provide insights into galaxy evolution include the baryonic Tully-Fisher relation (baryonic mass vs. rotational velocity; e.g. \citealt{mcgaugh00, lelli2019, ponomareva2018, ponomareva2021}); the relation between optical and \hi{} diameters \citep{BroeilsRhee1997,Leroy2008}; star formation histories, stellar masses and structural parameters \citep{Kauffmann2003}; and the mass-metallicity relation \citep{Tremonti2004}.

Another fundamental scaling relation for disc galaxies is the \hi{} size-mass relation. First discovered by \cite{BroeilsRhee1997}, it shows a tight correlation between the diameter of an \hi{} disc (\dhi), measured at a surface mass density level of 1 \solMassPcSquare{}, and its total enclosed \hi{} mass (\mhi). Recent studies have demonstrated that this relation also holds true over a wide range of galaxy types, such as large spirals \citep[][]{VerheijenSancisi2001,Swaters2002,Wang2013,Lelli2016,ponomareva2016}, late-type dwarf galaxies \citep[][]{Swaters2002,Begum2008,Lelli2016}, early-type spirals \citep[][]{Noordermeer2005}, irregulars \citep[][]{Lelli2016} and even for the ultra-diffuse galaxies (UDGs) discovered in the Arecibo Legacy Fast ALFA (ALFALFA) survey \citep[][]{Leisman2017,Gault2021}. Moreover, even though the intergalactic medium (IGM) of groups and clusters can affect the sizes of \hi\ discs due to ram-pressure and tidal interactions \citep{VerdesMontenegro2001}, it was shown that galaxies which reside in groups and clusters still follow the observed \hi{} size-mass relation as long as their discs are not too disrupted and the diameter can be traced out to 1 \solMassPcSquare{} \citep{VerheijenSancisi2001, chung2009}.

The galaxies in hydrodynamical simulations and semi-analytical models also follow the observed scaling relation between the \hi\ mass and size, with an analytically derived limit on its scatter of $\leq 0.1 ~\rm dex$ (e.g. \citealt{Wang2014,Marinacci2017,ElBaldry2018,Lutz2018}).
Although, environmental processes such as ram-pressure stripping may cause galaxy discs to truncate or have holes, this does not strongly affect the relation unless the disc is completely disturbed \citep{Stevens2019}. 
Moreover, \cite{Stevens2019} have shown that the robustness of the \hi\ size-mass relation makes it a valuable tool for theories of galaxy formation and evolution: the success of any model or simulation should be based on its ability to reproduce its scatter, slope and zero point with only a few percent uncertainty.

The largest observational work-to-date was undertaken by \cite{Wang2016}, who collected \hi\ sizes for more than 500 galaxies from 14 various projects, ranging over five decades in \mhi{}. They obtained a remarkably tight relation with a scatter of 0.06 dex (14\%). 
Although low-mass galaxies were found to have denser \hi\ discs than higher-mass galaxies due to their low angular momentum \citep{Lelli2016}, the tight power law correlation indicates a nearly constant characteristic \hi\ surface density within \dhi{} for most
galaxies \citep{Wang2016} -- regardless of their type, mass, or environment. This relation therefore suggests that all galaxies, from small dwarfs to large spirals, experience a similar process of evolution as long as they remain gas-rich.

Little is known about the resolved \hi\ content of galaxies located beyond the local Universe, mostly due to technical limitations such as lack of sensitivity and narrow frequency coverage of radio interferometers. To date, only a few surveys observed \hi\ in galaxies at $z>0.01$ (e.g., the Blind Ultra-Deep \hi\ Environmental Survey (BUDHIES; \citealt{gogate2020}) and HIGHz; \citealt{Catinella2015}). The highest redshift \hi{} detection until now is a starburst galaxy found in the COSMOS \hi{} Large Extragalactic Survey (CHILES) at $z = 0.376$ \citep[][]{Fernandez2016}. Therefore, the size-mass relation has only been studied for various nearby \hi{}-selected samples, and remains unexplored for large, homogeneous samples, which extend to higher redshifts. With the advent of deep \hi\ surveys with the Square Kilometre Array (SKA) pathfinders (and eventually the SKA itself), a new window is opening for studying the \hi\ content of galaxies beyond the local volume. 

This work is based on the Early Science data from the MeerKAT International GHz Tiered Extragalactic Exploration (MIGHTEE) survey
\citep{Jarvis2016}. Our sample comprises 276 galaxies detected as part of the spectral line component of MIGHTEE. The sample spans more than four orders of magnitude in \hi\ mass and one billion years in lookback time ($z \leq 0.084$). Therefore, we are able to study the \hi\ size-mass relation for the first time beyond the local Universe using a homogeneous data set, and explore its possible evolution with redshift.

Our paper is structured as follows: we summarize the MeerKAT observations and data reduction strategy in Section \ref{sec:obs}. Sample selection and morphological galaxy classification are described in Section \ref{sec:sample}. In Sections \ref{sec:HI_size} and \ref{sec:HI_mass}, we present the measurements of the \hi\ size and mass of our sample galaxies, respectively. We analyse our results and compare them with existing studies in Section \ref{sec:results}. Section \ref{sec:conclusion} summarizes our findings. 

Throughout this paper, we assume $\Lambda$CDM cosmology parameters of $\Omega_{\rm m} = 0.3$, $\Omega_{\Lambda} = 0.7$ and $H_0 = 70$ \kms{}\,$\rm Mpc^{-1}$, for ease of comparison with previous results.

\section{MEERKAT observations and data reduction}\label{sec:obs}

MIGHTEE is one of the eight Large Survey Project (LSP) of MeerKAT \citep{Jonas2016}. MIGHTEE science cases include studies of the radio continuum (see \citealt{Heywood2021} for the early science data release), \hi{} in emission (MIGHTEE-\hi{}), \hi\ absorption and polarisation in galaxies. This study focuses on the MIGHTEE-\hi\ part of the survey, and uses the Early Science data, details of which can be found in \cite{Maddox2021}.

The observations were carried out with the full MeerKAT array between mid-2018 and mid-2019 in the L-band. The data were collected using the 4k spectral line correlator mode, with a channel width of 209 kHz, which corresponds to a velocity resolution of 44.11 \kms{} at $z = 0$.  The observations were carried out over two of four MIGHTEE fields described in \cite{Jarvis2016}, covering approximately 3.5 deg$^{2}$ of the sky in XMMLSS and 1.5 deg$^{2}$ in COSMOS, resulting in a total area of ${\sim} 5 ~\rm deg^2$. 
\begin{figure*}
	\includegraphics[width=\linewidth]{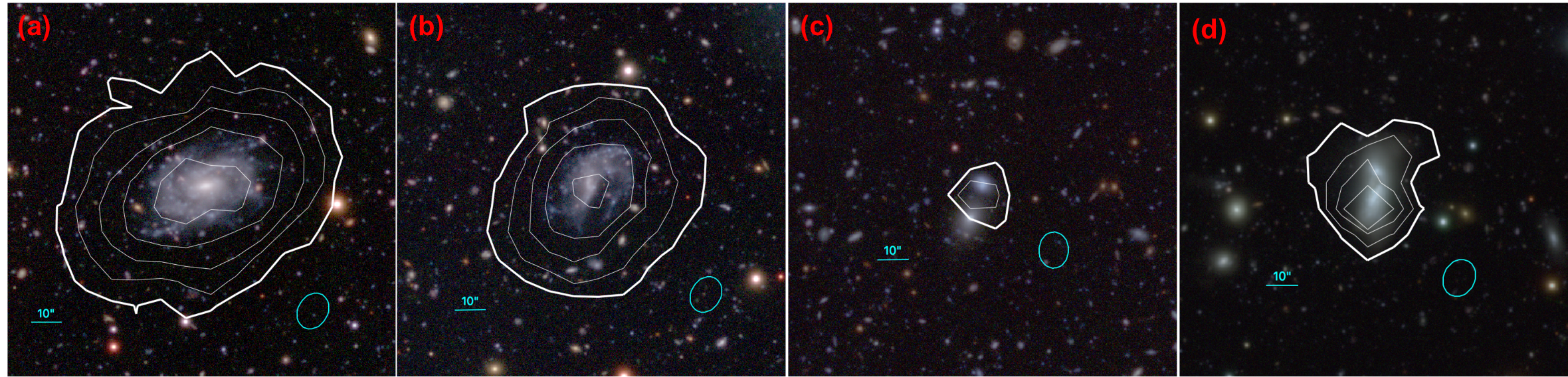}
	\centering
	\caption{Examples of galaxies of different morphological types detected within MIGHTEE-\hi{}. Panel (a): an undisturbed spiral galaxy at $z=0.033$; panel (b): an irregular galaxy at $z=0.044$; panel (c): an early-type galaxy (the northern counterpart) at $z=0.037$; panel (d): a merging system at $z=0.032$. The optical cutouts are from the HSC \emph{gri} images, with the \hi{} column density contours overlaid in white. The $1.25\times 10^{20}$ atoms\pcm\ (1 \solMassPcSquare{}) contour is shown in bold. Thin contours correspond to $3, 6, 10, 15 \times 10^{20}$ atoms\pcm\ for panels (a) and (b), $2\times 10^{20}$ atoms\pcm\ for panel (c), and $2, 3, 4 \times 10^{20}$ atoms\pcm\ for panel (d). A 10\arcsec\ scale bar and the synthesized beam size are displayed in each panel, in which panel (c) has a slightly different angle due to it being from a different field than the other panels.}
	\label{opt}
\end{figure*}

Data calibration tasks such as flagging, delay, bandpass, gain and self-calibration were done with the \textsc{ProcessMeerKAT}\footnote{https://idia-pipelines.github.io} pipeline Collier et al. (in preparation). 
This is a \textsc{Casa}\footnote{http://casa.nrao.edu}-based pipeline developed at the Inter-University Institute for Data Intensive Astronomy (IDIA)\footnote{https://idia.ac.za}. 

Visibility based continuum subtraction was done in two steps -- an initial subtraction of the best clean component continuum model and the subtraction of a polynomial fit to the per-baseline/per-integration spectrum. The residual visibilities were imaged using \textsc{Casa}'s \textsc{TCLEAN} task with \cite{Briggs1995} weighting \textsc{(robust = 0.5)}. A final step of median-filtering was done on the cubes to reduce the impact of continuum-subtraction errors. Full details of the data processing can be found in Frank et al. (in prep).

The \hi{} cube is ${\sim} 2.3^\circ \times 2.3^\circ$ per pointing. The dirty beam FWHM's are $\ang{;;14.5} \times \ang{;;11}$ and $\ang{;;12} \times \ang{;;10}$ for the COSMOS and XMMLSS fields, respectively.
A 3$\sigma$ column density sensitivity in COSMOS is $4.05 \times 10^{19}$ atoms\pcm\, and $9.83 \times 10^{19}$ atoms\pcm\ in XMMLSS.
The observational and imaging parameters of the Early Science data are summarized in Table \ref{tab1}.

\begin{table}
	\small
	\centering
	\caption{A summary of the MeerKAT observing parameters and imaging of the MIGHTEE-\hi\ Early Science data.}
	\label{tab1}
	\begin{tabular}{l l}
		\hline
		\hline
		Observing parameters & Value \\
		\hline
		Survey area & 1.5 deg$^{2}$ (COSMOS field) \\
		& 3 $\times$ 1.2 deg$^{2}$ (XMMLSS field) \\
		Total integration time	& ${\sim}17$ hrs (COSMOS) \\
		& 3 $\times$ 13 hrs (XMMLSS)\\
		Spectral resolution &	209 kHz
		\\
		Velocity resolution &	44.11 \kms{} at $z = 0$ \\
		Velocity range & 86 -- 24205 \kms{}  \\
		PSF (FWHM) & $\ang{;;14.5} \times \ang{;;11}$ (COSMOS)\\
		& $\ang{;;12} \times \ang{;;10}$ (XMMLSS)\\
		Pixel / Image size & $\ang{;;2}$ / $4096 \times 4096$ \\
		$3\sigma$ \hi\ column density sensitivity & $4.05 \times 10^{19}$ atoms\pcm\ (COSMOS) \\
		& $9.83 \times 10^{19}$ atoms\pcm\ (XMMLSS)\\
		\hline
	\end{tabular}
\end{table}

\section{Sample selection}\label{sec:sample}

Source finding was performed visually on the Early Science data cubes covering 1310\,--\,1420\,MHz, as described in \cite{Maddox2021}. This resulted in 276 unique \hi{} detections, forming the basis for our analysis. There were no restrictions on redshift, morphology, mass, or environment, i.e. all detections were initially considered for the current study. 

After an in-depth examination of the sources, four galaxies were removed from the sample because they were classified as intermediate-stage mergers, i.e. when a system comprises two distinct stellar discs, but a single \hi{} structure encompassing both galaxies. Early-stage mergers, where \hi\ discs are still associated with the individual galaxies were kept in the sample. Late-stage mergers, where the \hi\ and the stars have both merged into a single structure were also retained. After removing these four galaxies, our sample was reduced to 272 objects. 

We further imposed that galaxies must be resolved with at least one and a half resolution elements across the major axis, and that the radial extent of the inclination corrected surface mass density ($\Sigma_{\rm HI}$, see Section \ref{sec:HI_size}) must reach 1 \solMassPcSquare{} ($1.249 \times 10^{20}\ \rm atoms\ cm^{-2}$). Even though the measurement of $D_{\rm HI}$ at 1 \solMassPcSquare{} is a subjective choice, \citet{Wang2016} has shown that a diameter $D_{\rm HI}$ defined at 1 \solMassPcSquare{} encloses most of the \hi{} mass of a galaxy and is measurable for most of the galaxies, even for small \hi{} discs that are close to being unresolved. 

Additionally, the  $\Sigma_{\rm HI}$ contour at 1 \solMassPcSquare{} should not be strongly disrupted due to ongoing mergers or tidal interactions (for details, see Section \ref{sec:HI_size}). A total of 204 out of 272 galaxies satisfy our selection criteria, and form the final sample for our study of the  $D_{\rm HI}$\,--\,$M_{\rm HI}$ relation. The median resolution of our resulting sample is three beams across the major axis, with only 6 galaxies being resolved with more than 10 beams.

\subsection{Morphological classification}\label{sec:morpho}

The galaxies in our sample were visually classified based on their optical morphology. Three of the authors (SR, AP and NM) inspected three-colour images created from Subaru HyperSuprimeCam $g, r,$ and $i$-band images (HSC; \citealt{Aihara2018}, \citeyear{Aihara2019}). For the few (16 of 204) objects lying outside the HSC imaging footprint, the three-colour images from the Sloan Digital Sky Survey Data Release 16 (SDSS DR16; \citealt{Ahumada2020}) were used. The SDSS imaging is substantially shallower than that from HSC, but the number of objects is small, and the different imaging does not affect our results. The majority vote of the three classifiers was taken as the adopted morphology. While automated morphological classification algorithms are used for large datasets (e.g. the Zurich Estimator of Structural Type, \citealt{Scarlata2007}), visual classification is still in use \citep[e.g.][]{Hashemizadeh2021}.
The \hi\ morphology was not used for the classification except to remove merging systems as noted before.

The galaxies were divided into four morphological categories: spirals (SP), early-types (ET), irregulars (IR) and mergers (ME). No distinction was made between irregulars and dwarf irregulars as no stellar mass information was used in the classification. ET galaxies have smooth, centrally concentrated morphology, whereas SP galaxies show clear spiral arms originating from either a central bulge or bulge/bar. IR objects show no regular patterns, and ME systems show signs of interaction between two or more galaxies, including disturbed morphology or tidal streams.
Of the 204 galaxies in the original sample, there are 148 SP, 40 IR, 12 ME and 4 ET. Examples of the four classes are given in Fig.~\ref{opt}.

\begin{figure*}
	\centering
	\includegraphics[width=1.0\linewidth]{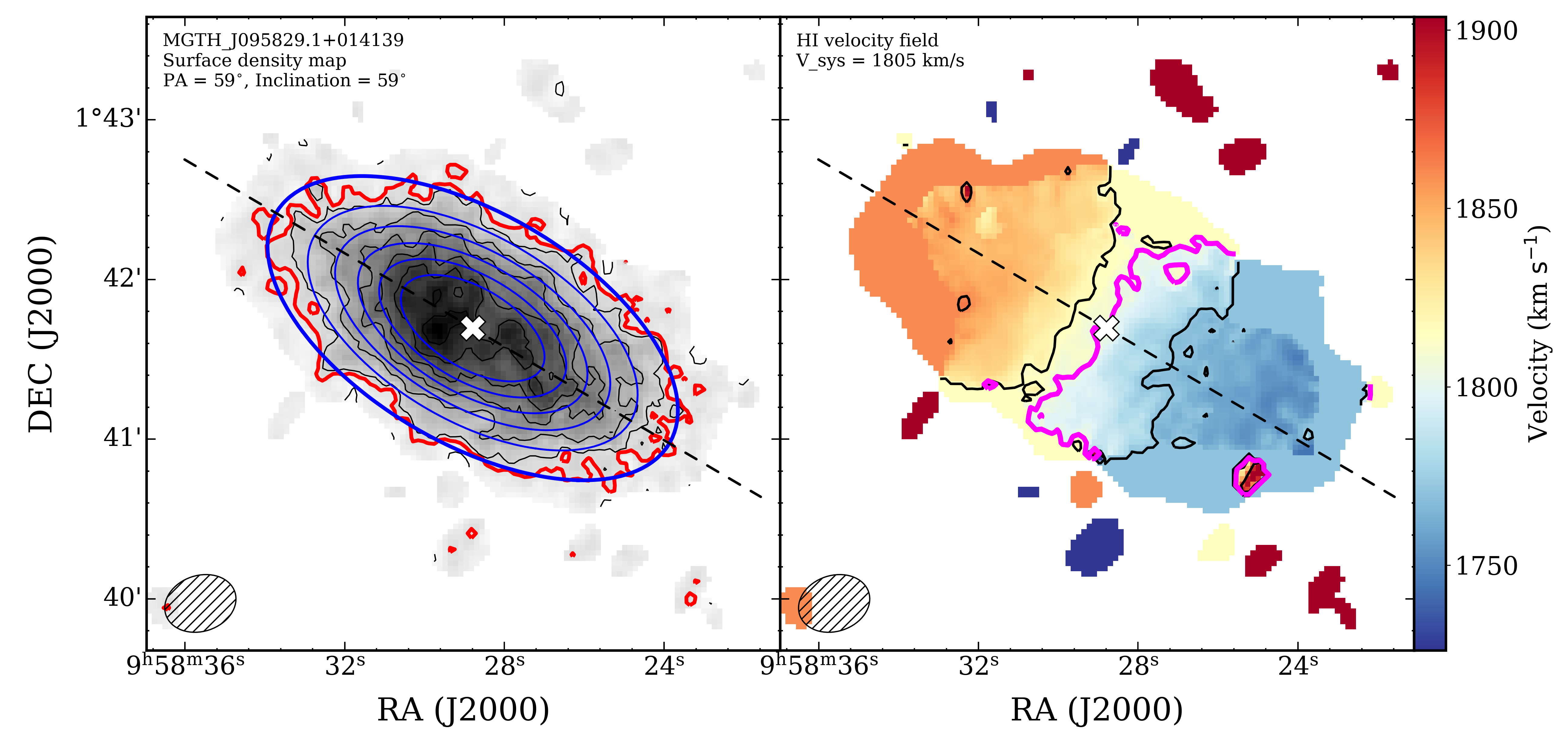}
	\centering
	\caption{Example illustrating the 2D Gaussian fitting procedure (source MGTH\_J095829.1+014139). Left panel: Surface mass density map (grey scale and black thin contours). The contour levels are 10.5, 9, 7.5, 6, 4.5, 3, 1.5 \solMassPcSquare{} and the 1 \solMassPcSquare{} contour is highlighted in red. The fitted 2D Gaussian function is shown by the thin blue ellipses; the fitted $D_{\rm HI}$ is highlighted by the thick blue outermost ellipse. Right panel: velocity field (moment 1). Iso-velocity contours are shown in black and separated by 45 \kms{}. The measured systemic velocity is indicated by the thick magenta line. In both panels, the white cross marks the fitted ellipse centre, the dashed line represents the fitted major axis of the ellipse from the 2D Gaussian fitting. The beam size is indicated in the lower left corner of each panel.}
	\label{fig:2d_fit}
\end{figure*}

\section{\HI{} size}\label{sec:HI_size}

In order to measure the size of the \hi{} discs, we use the moment 0 maps, produced for each detection as described in detail in \citet{ponomareva2021} and \citet{Ranchod2021}. 

We convert moment 0 maps from the units of flux density to surface mass density following the prescription from \citet{Meyer2017}:
\begin{equation}
\left(\frac{\Sigma_{\mathrm{HI}}}{\mathrm{M_{\odot}\,pc^{-2}}}\right) = 1.00 \times 10^{4} (1+z)^3 \left( \frac{S}{\rm Jy\,km\,s^{-1}} \right) \left( \frac{\Omega_{\rm bm}}{\rm arcsec^{2}} \right)^{-1},
\label{eq:surf_dens}
\end{equation}
where $z$ is redshift, $S$ is flux density and  $\Omega_{\rm bm} = \frac{\pi b_{\rm maj} b_{\rm min}}{4 \rm ln(2)}$ is the solid angle of the synthesized beam with major axis $b_{\rm maj}$ and minor axis $b_{\rm min}$.  

Once the surface mass density maps are obtained, we use the following approach to determine $D_{\rm HI}$:
\begin{enumerate}[wide, labelwidth=!, labelindent=4pt]
	\item We use a 2D elliptical Gaussian function to fit the surface mass density map and measure its diameter at the 1 \solMassPcSquare{} contour level (see left panel of Fig. \ref{fig:2d_fit}). It is important to note that the \hi{} radial distribution is not Gaussian, and \hi{} radial profiles often reveal a depletion at the center \citep[e.g.][]{Wang2014,Martinsson2016}. However, we are only interested in \hi{} distribution at the outer part of the \hi{} disk where the \hi{} diameter is measured. In addition, the majority of our sample galaxies are only marginally resolved with a median resolution of 3 beams. Therefore, a 2D elliptical Gaussian function works similar to a simple ellipse fitting. The function is expressed as: 
	\begin{equation}
	f(x,y) = A \exp{(-(a(x-x_0)^2 +2b(x-x_0)(y-y_0) + c(y-y_0)^2))},
	\label{2d_gauss}
	\end{equation}
	where $A$ is the amplitude of the Gaussian peak in \solMassPcSquare{}, $(x_0,y_0)$ its centre position in pixels, and $a,b,$ and $c$ are defined as:
	
	\begin{align}
	a = \frac{\cos^2{\theta}}{2\sigma_X^2} + \frac{\sin^2{\theta}}{2\sigma_Y^2},\\
	b = -\frac{\sin{2\theta}}{4\sigma_X^2} + \frac{\sin{2\theta}}{4\sigma_Y^2},\\
	c = \frac{\sin^2{\theta}}{2\sigma_X^2} + \frac{\cos^2{\theta}}{2\sigma_Y^2}.
	\end{align}
	In the above equations, $\theta$ is the position angle in radians, $\sigma_X$ and $\sigma_Y$ are the semi-major and semi-minor axes of the disk.
	We assume the following initial values for the parameters $A, x_0, y_0,\sigma_X, \sigma_Y$ and $\theta$ to optimize the fitting process:
	\begin{itemize}[noitemsep,topsep=-8pt]
		\item[--] the amplitude $A$ is set to the maximum pixel value in the map in units of \solMassPcSquare{},
		\item[--] the estimated \hi{} emission centre $x_0, y_0$ is assumed to be at the centre of the map i.e. half the number of pixels contained in the x and y axis of the map,
		\item[--] $\sigma_X$ and $\sigma_Y$ are set to 10 pixels as 20 pixels ($\ang{;;40}$) is the typical extent of a source in the MIGHTEE-\hi\ early science data,
		\item[--] the position angle $\theta$ was assumed to be 0. 
	\end{itemize}
	\item With these estimates, we perform a non-linear least square fitting and obtain optimal values for the 2D Gaussian parameters listed in step (i). We use the 1 \solMassPcSquare{} ellipse to obtain the corresponding best-fitting central position of the \hi{} emission $(x_0,y_0)$, the major and minor axis values, and the position angle of the ellipse. 
	
	As an example of the 2D Gaussian fitting, Fig. \ref{fig:2d_fit} displays the fitting outcome for a galaxy with $\rm V_{sys} \sim 1\,805 ~\rm km\,s^{-1}$ ($z = 0.006$), where ellipses, resulted from the fit are over-plotted on top of the surface mass density map (left panel). The velocity field (right panel), though not being used in the fitting process, is shown to demonstrate that the fitted centre and major axis coincide with the kinematic centre and major axis.

	To derive the inclination angle of the plane of the galaxy in degrees, we use:
	\begin{equation}
	\cos^2{(i_{\rm HI})} = \frac{(2.355 \, \sigma_{Y})^2 - b^2_{\rm min}}{(2.355 \, \sigma_{X})^2 - b^2_{\rm maj}},
	\label{eq:inc}
	\end{equation}
	where $b_{\rm maj}$ and $b_{\rm min}$ are the major and minor axes of the synthesized beam \citep{VerheijenSancisi2001}. Galaxies with low spatial resolution might appear rounder due to the beam smearing. The inclusion of the synthesized beam in Eq. \ref{eq:inc} will account for the beam smearing effect.  However, it is important to note that this correction assumes that the axes of the beam are aligned with the axes of the galaxy, which is not necessary the case for our data. We ran the tests to evaluate the additional error on the inclination caused by possible misalignment of the beam and the source axes. We found that for marginally resolved galaxies ($<3$ beams) the  error on inclination is ${\sim} 7$ degrees, for galaxies resolved with 3\,-\,5 beams the error is ${\sim} 3$ degrees, while it is negligible for galaxies resolved over more than 5 beams. We add these errors in quadrature to the inclination measurement error of ${\sim} 5$ degrees \citep{ponomareva2021}. This results in the total error on inclination for marginally resolved galaxies : ${\sim} 8.5$ degrees, for galaxies resolved with 3\,-\,5 beams: 5.8 degrees, and 5 degrees for galaxies resolved with more than 5 beams. We also note that our inclinations, measured from the \hi{} maps are in an excellent agreement with inclinations measured from the optical photometry and with inclinations derived with the 3D kinematic modelling for a sub-sample of 67 galaxies from \cite{ponomareva2021}.
	\item We correct each surface mass density map for the measured inclination by multiplying each unmasked pixel value of the map by the cosine of the inclination angle. Then, we repeat steps (i) and (ii). 
	\item $D_{\rm HI}$ (2$\sigma_X$) is then  measured from the inclination corrected maps along the major axis of the best-fitting ellipse corresponding to the surface mass density contour of 1 \solMassPcSquare{}.
	\item To complete the \hi{} size measurements, we correct $D_{\rm HI}$ for beam smearing effect using the following prescription from \cite{Wang2016}:
	\begin{equation}
	D_{\rm HI,corr} = \sqrt{D^2_{\rm HI} - b_{\rm maj} \times b_{\rm min}},
	\label{eq:corr_smear}
	\end{equation}
	where $D_{\rm HI,corr}$ is the corrected \hi\ size which we use to construct the $D_{\rm HI}$\,--\,$M_{\rm HI}$ relation. This correction removes a systematic bias which induces an over-estimation of $D_{\rm HI}$ for marginally resolved galaxies. 
	
	The conservative uncertainty on $D_{\rm HI,corr}$ is assigned as half of the synthesized beam major axis $b_{\rm maj}$, expressed in kpc, and includes the uncertainty on the inclination angle. The error on the cosmological luminosity distance ($D_L$) is also propagated during the conversion. The latter was derived by adopting the channel width as the error on the systemic velocity and 2.2 $\rm km~s^{-1}~Mpc^{-1}$ as the error on the Hubble constant \citep[][]{Hinshaw2013}. The resulting error on $D_L$ is ${\sim} 3.5$ Mpc. The uncertainty on $\log D_{\rm HI}$ slightly increases with distance and is ${\sim}$0.02 kpc at $z = 0$ and ${\sim}$0.11 kpc at $z \simeq 0.084$.
\end{enumerate}

\section{\HI{} mass}\label{sec:HI_mass}
To construct the $D_{\rm HI}$\,--\,$M_{\rm HI}$ relation, we measure the total \hi{} mass enclosed within the moment 0 map of a galaxy, since the amount of \hi\ beyond the diameter at 1 \solMassPcSquare{} is negligible for our sample.
By assuming an optically thin gas ($\tau \ll 1$) with no significant self-absorption, we use the following equation from \cite{Meyer2017} which uses the cosmological luminosity distance $D_L$ to the galaxy to determine the \hi\ mass:

\begin{equation}
\left(\frac{M_{\rm HI}}{M_{\odot}}\right) \simeq \frac{2.356 \times 10^5}{(1+z)} \left( \frac{D_L}{\rm Mpc} \right)^2 \left( \frac{S}{\rm Jy\,km\,s^{-1}} \right),
\label{himass}
\end{equation}
where $z$ is the redshift, and $S = \int S_v dv$ is the integrated flux density derived from the moment 0 map. \\

The uncertainty on the integrated flux $S$ was estimated from the mean RMS noise within four emission-free regions around the detection \citep{mpati2016, ponomareva2021}. 
As a result, an \hi\ mass uncertainty of ${\sim} 5\%$ is measured for large mass galaxies ($M_{\rm HI} > 10^9 ~\rm M_{\odot}$), ${\sim} 10\%$ for $10^8 < M_{\rm HI} < 10^9 ~\rm M_{\odot}$ while the error can reach up to 20\% for galaxies below $M_{\rm HI} < 10^8 ~\rm M_{\odot}$ (see \citealt{Maddox2021} for details).

\section{Results}\label{sec:results}
This section presents the correlation between the \hi\ mass and the \hi\ size of the sample of 204 galaxies. We also show the comparison between the resulting $D_{\rm HI}$\,--\,$M_{\rm HI}$ relation and previous studies at $z = 0$, as well as its evolution as a function of redshift.

\subsection{\dhi\,--\,\mhi\ relation}
\label{sec:the_relation}
We construct the \dhi\,--\,\mhi\ relation and study its statistical properties, such as slope, zero point and scatter, by performing a power-law fit to the relation with a maximum likelihood function that takes measurement errors of both parameters into account, and assumes a Gaussian intrinsic scatter along the vertical direction to the best-fitting line (see Eq. A4 in \citealt{lelli2019}). We can therefore investigate for the first time whether the previously reported small scatter of the relation ($\sigma=0.06$ dex, \citealt{Wang2016}) is due to measurement errors or is an intrinsic property.

We use the standard affine-invariant ensemble sampler for Markov chain Monte Carlo (MCMC) {\it emcee}\footnote{https://emcee.readthedocs.io/en/stable/} \citep{Foreman2013} to map the posterior distributions of the main statistical properties: slope, zero point and intrinsic scatter of the relation, following the prescriptions described in \citet{lelli2019}. 

For the fit, we initialize the chains with 50 random walkers, run 1000 iterations and re-run the simulation with 1000 steps. The starting position of the walkers is set randomly within realistic prior ranges: slope [0.1, 1], zero point [-6, 0] and intrinsic vertical scatter ($\sigma_{\text{int}}$) [0.01, 0.1]. The convergence of the chains is then checked visually.

\begin{figure}
	\centering
	\includegraphics[width=\linewidth]{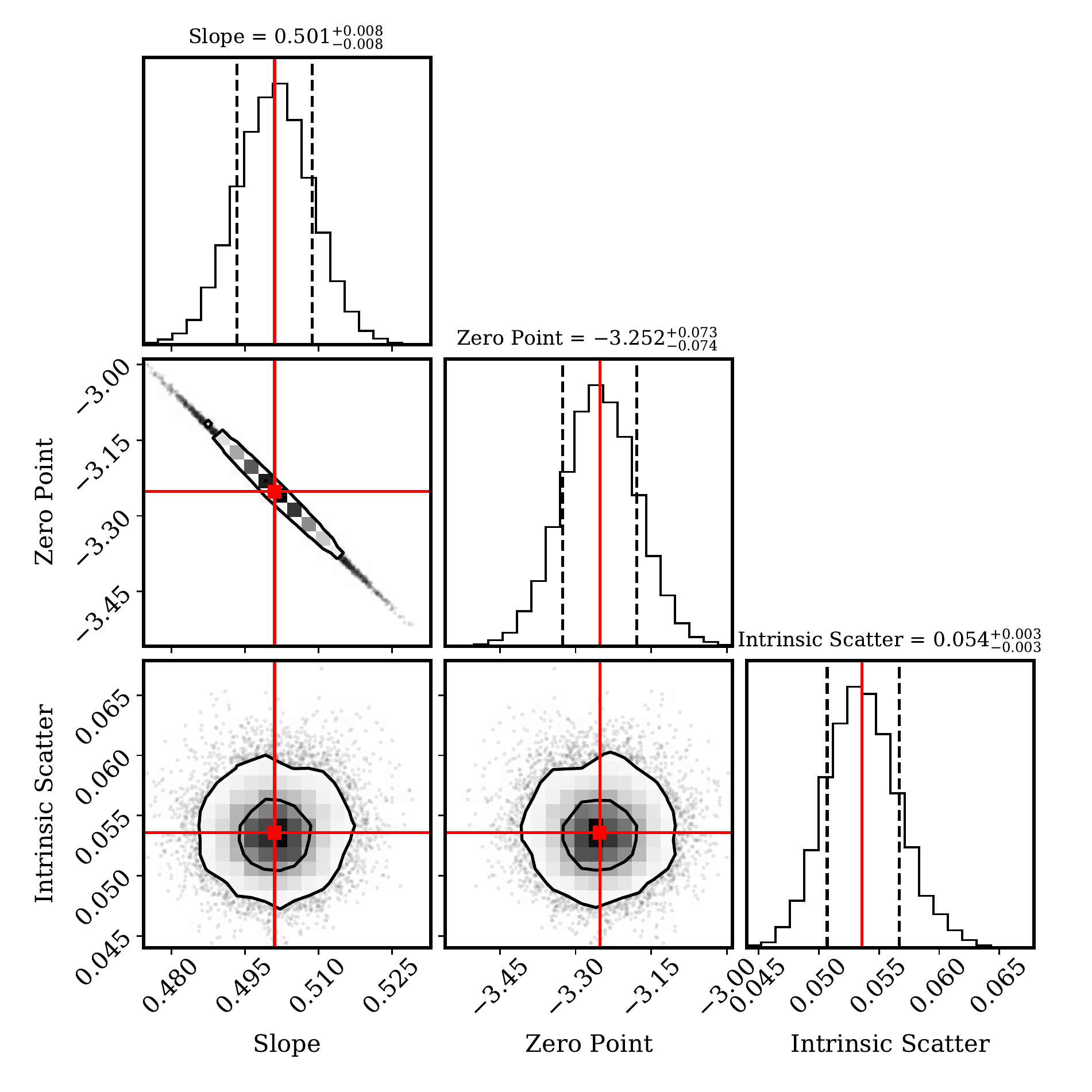}
	\centering
	\caption{The posterior distributions of the slope, zero point and the intrinsic scatter of the $D_{\rm HI}$\,--\,$M_{\rm HI}$. The best-fitting median values are indicated with the red squares and solid lines. Black contours are 68 ($1\sigma$) and 95 ($2\sigma$) per cent confidence regions.}
	\label{fig:bayes}
\end{figure}

\begin{figure*}
	\centering
	\includegraphics[width=0.85\linewidth]{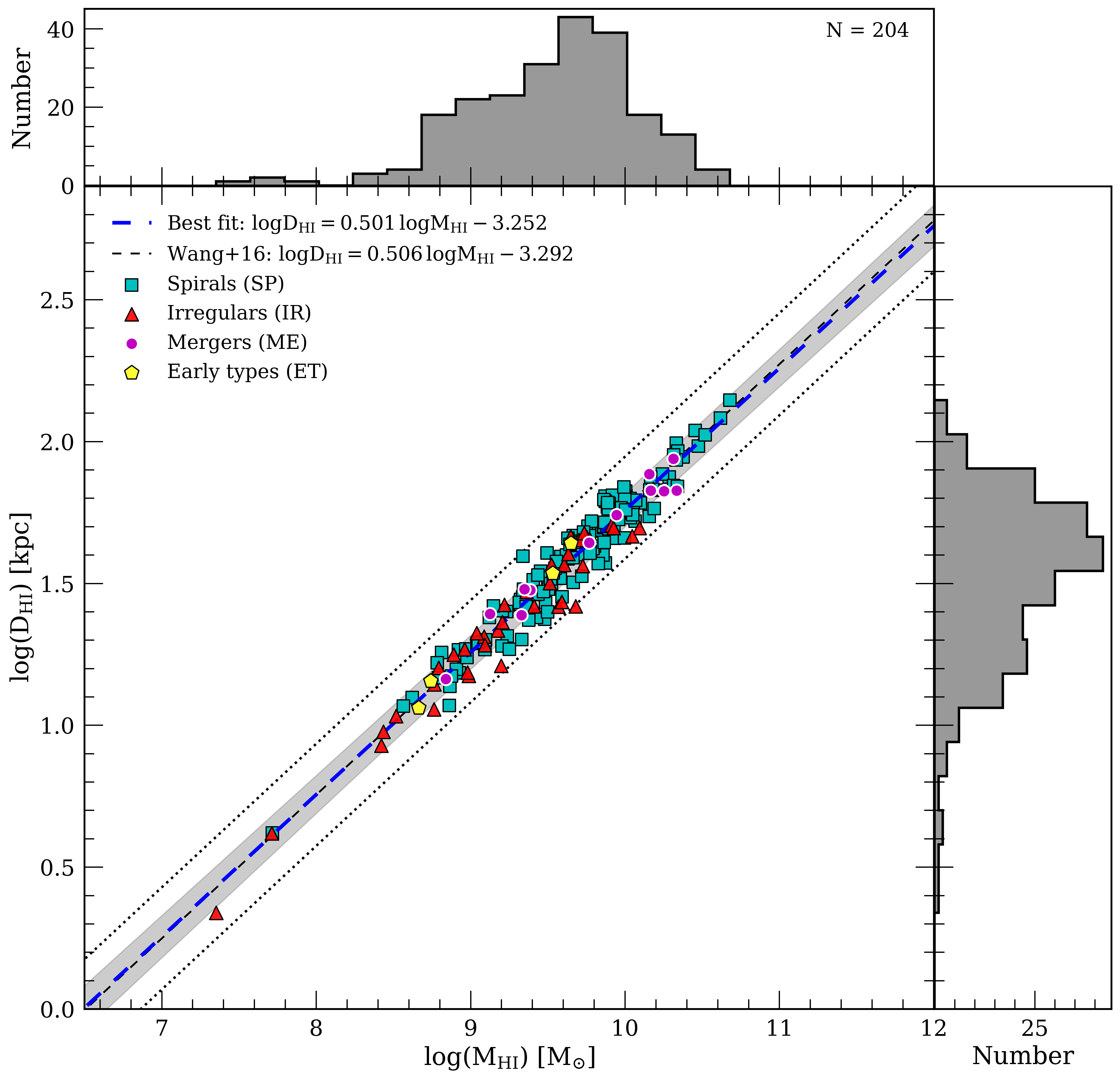}
	\centering
	\caption{\dhi\,--\,\mhi\ relation for 204 inclination-corrected galaxies. 
		Each symbol corresponds to one of the four morphological classifications: 148 spiral galaxies (SP: cyan squares), 40 irregular galaxies (IR: red triangles), 4 early-type galaxies (ET: yellow pentagons), and 12 merging systems (ME: magenta circles). The best-fitting relation is shown with the thick dashed blue line, while the best fit from \citet{Wang2016} is shown with a dotted black line. The shaded region indicates the $1\sigma$ uncertainty from the MCMC posteriors (see Fig. \ref{fig:bayes}) and the black dotted lines delimit the $3\sigma$ scatter from \citet{Wang2016} 
		relation. The histograms on the sides display the \hi\ mass and \hi\ size distributions of the galaxies in the sample, respectively.}
	\label{fig:D_HI_M_HI:morpho}
\end{figure*}

The posterior distributions of the parameters are shown in Fig. \ref{fig:bayes} and their median values are listed in Table \ref{tab:bayesian_results}. 
The resulting $D_{\rm HI}$\,--\,$M_{\rm HI}$ relation for our data and the associated 1$\sigma$ uncertainty from the MCMC posterior distribution is presented in Fig. \ref{fig:D_HI_M_HI:morpho}. 
We find the relation to be:
\begin{equation}
\log D_{\rm HI} = 0.501^{+0.008}_{-0.008}~\log M_{\rm HI} - 3.252^{+0.073}_{-0.074}.
\label{eq:hi_size_mass_mightee}
\end{equation}

The non-zero intrinsic scatter ($\sigma_{int}=0.054 \pm 0.003$ dex) is comparable to the observed scatter ($\sigma=$ 0.057 dex) and suggests that the scatter of the relation cannot be explained just with the measurement errors, but rather is an intrinsic property of the relation allowing for a ${\sim} 10\%$ variation of the $D_{\rm HI}$ at a fixed \hi{} mass. To assess whether the intrinsic scatter is introduced by an underestimation of the measurement errors, we have repeated the fit using measurement errors which are 2, 3, and 4 times larger than the original values in both directions. Consequently, all resulting intrinsic scatters were non-zero.

Furthermore, our relation is in excellent agreement with \cite{Wang2016}, who found a slope of $0.506\pm0.003$, an intercept $-3.293\pm0.009$, and an observed scatter of $\sigma = 0.06$ dex. 

Fig. \ref{fig:D_HI_M_HI:morpho} shows the distribution of the detections in the relation with respect to the $1\sigma$ confidence region of the MCMC fit and the $3 \sigma$ scatter of the \cite{Wang2016} relation. It is observed that all detections lie within the black dotted lines which are the $3\sigma$ scatter of the \cite{Wang2016} relation.
\begin{figure*}
	\centering
	\includegraphics[width=0.8\linewidth]{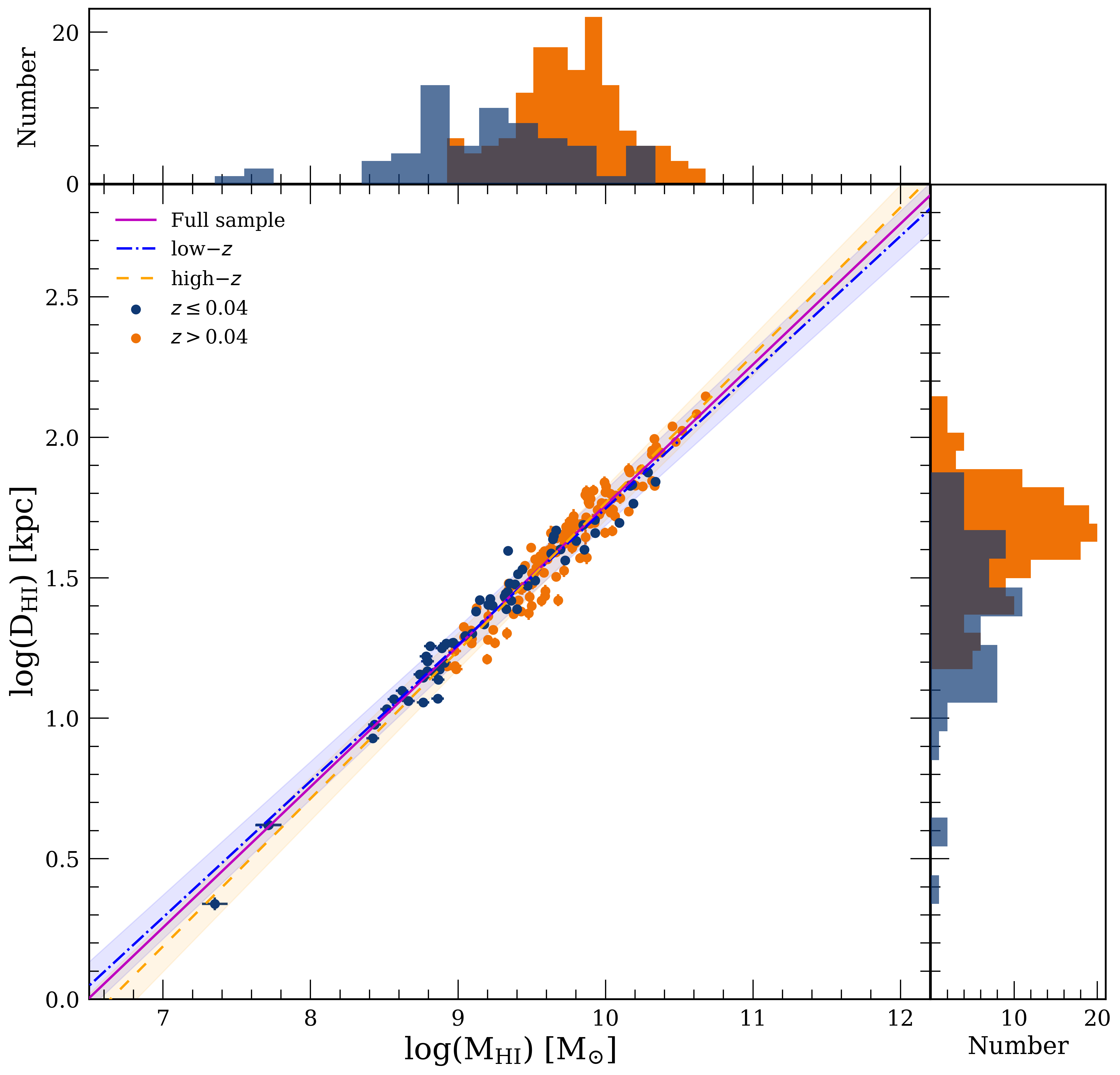}
	\centering
	\caption{$D_{\rm HI}$\,--\,$M_{\rm HI}$ relation of galaxies in the MIGHTEE-\hi{} survey for two redshift bins. The errorbars on the points are approximately the same size as the points themselves. The high-redshift subsample ($z > 0.04$) is presented in orange circles, the low-redshift subsample ($z \leq 0.04$) in blue. The magenta solid line indicates the best fit from the maximum-likelihood method of the full sample, while the orange dashed and the blue dashed-dotted lines represent the best fits from the high-redshit and the low-redshift subsamples respectively. Their respective 1$\sigma$ scatter is indicated by the shaded orange and blue regions. The \hi\ mass and \hi\ size distributions for the two redshift bins are displayed on the upper and right sides of the main frame.}
	\label{fig:D_HI_M_HI:redshift}
\end{figure*}

\subsection{$D_{\rm HI}$\,--\,$M_{\rm HI}$ relation and galaxy type}
\label{sec:relation and morphology}
Previous studies have focused on targeted morphologies of galaxies (e.g. \citealt{Begum2008} with dwarf galaxies, \citealt{Noordermeer2005} with early-type galaxies). The absence of a homogeneous large sample containing various types of galaxies has led to the study of the relation from compilations of data from various surveys \citep{Lelli2016,Wang2016}. This work is based on a ``blind'' survey, and as such, is not morphologically selected. To obtain further insight into the $D_{\rm HI}$\,--\,$M_{\rm HI}$ relation, we explore the morphological properties of our sources, as classified in Section \ref{sec:morpho}. The dominant morphological type in our sample is very similar to \cite{Wang2016} (SP and IR, with a few ET). Fig. \ref{fig:D_HI_M_HI:morpho} shows the $D_{\rm HI}$\,--\,$M_{\rm HI}$ for our sample galaxies, grouped by morphological type, revealing that each subset of morphological types all lie on the relation and exhibits a low intrinsic scatter. The majority of SP/IR lie within the $1-2\sigma$ confidence region of the relation. It is important to note that our sample is \hi{}-selected, and thus is sensitive towards relatively \hi{}-rich galaxies. This is in contrast to the \hi{} survey of early-type galaxies ATLAS$^{\rm 3D}$ \citep[][]{Serra2012} which detected \hi{}-poor ET galaxies below our column density limit and found that the typical \hi{} column density of ETs is lower than of spiral galaxies (see Fig. 10 in \citealt{Serra2012}). Consequently, as far as ETs are concerned in this relation, we could not detect the faintest and \hi{}-poorest ones due to a selection effect.

\begin{table}
	\centering
	\caption{Median values of the posterior distributions of the MCMC-based linear fit.  Median \hi\ mass and \hi\ sizes for the $D_{\rm HI}$\,--\,$M_{\rm HI}$ relation at different redshift bins, and the corresponding $\log D_{\rm HI}$ for a fixed $\log M_{\rm HI}$.}
	\begin{tabular}{l c c c}
		\hline
		\hline
		& $0 \leq z \leq 0.084$ & $ z \leq 0.04$ & $ z > 0.04$\\
		\hline
		Sample   & N = 204  & N = 63 & N = 141\\ [0.04cm]
		Slope   & $0.501^{+0.008}_{-0.008}$ & $0.485^{+0.012}_{-0.011}$& $0.526^{+0.013}_{-0.013}$\\ [0.1cm]
		Intercept  & $-3.252^{+0.073}_{-0.074}$ & $-3.104^{+0.105}_{-0.107}$ & $-3.494^{+0.125}_{-0.128}$\\ [0.08cm]
		Scatter ($\sigma$) & 0.057 & 0.054 & 0.058 \\ [0.04cm]
		Intrinsic scatter ($\sigma_{\rm int}$) & $0.054^{+0.003}_{-0.003}$ & $0.052^{+0.006}_{-0.005}$ & $0.053^{+0.004}_{-0.004}$\\ [0.1cm]
		Median $\log(M_{\rm HI}[\rm M_{\odot}])$ & 9.64 & 9.22 & 9.75\\
		Median $\log(D_{\rm HI}[\rm kpc])$ & 1.59 & 1.40 & 1.65\\ [0.08cm]
		$\log D_{\rm HI}$ & $1.579^{+0.015}_{-0.017}$ & $1.573^{+0.021}_{-0.028}$ & $1.579^{+0.029}_{-0.024}$ \\
		(at $ \log M_{\rm HI} =9.64\,\rm M_{\odot}$) & & & \\
		\hline
	\end{tabular}
	\label{tab:bayesian_results}
\end{table}

We assess the intrinsic scatter and slope of the relation as a function of morphology. The 148 spiral galaxies show a tight intrinsic scatter of $0.053^{+0.004}_{-0.003}$ with a slope of $0.491\pm 0.010$, representing galaxies with large and well-defined discs. For the 40 irregulars, we find an intrinsic scatter of $0.061^{+0.009}_{-0.007}$ and a slope of $0.492^{+0.017}_{-0.016}$. This result indicates that the slopes of the $D_{\rm HI}$\,--\,$M_{\rm HI}$ relations for SP and IR are statistically similar, and their intrinsic scatters are consistent within 2$\sigma$ error.

As highlighted in Section \ref{sec:sample}, only early and late-stage mergers, where the \hi{} disc belongs to a single galaxy, were included in the ME sample. This would explain why the MEs in our sample lie on the relation. Due to the small sample sizes, we could not investigate the intrinsic scatter of MEs and ETs separately.

\subsection{$D_{\rm HI}$\,--\,$M_{\rm HI}$ relation and environment}\label{sec:relation and environment}

The \hi\ content of galaxies is known to be sensitive to the environment \citep{Haynes1984}. \cite{VerdesMontenegro2001} have shown that the sizes of the \hi\ discs of galaxies in group environments are influenced by tidal interactions. Continuous tidal stripping due to the IGM in groups can lead to the perturbed \hi\ discs and \hi\ deficiencies in galaxies. This was also investigated for simulated galaxies by \cite{Stevens2019}, who showed that environmental processes, such as ram-pressure stripping, may cause disc truncation.

As mentioned in Section \ref{sec:sample}, we did not impose any environment-based constraint on our sample. There are indeed a variety of large-scale structures detected within the Early Science volume -- the most prominent of which is the large galaxy group at $z \sim 0.044$, recently presented by \cite{Ranchod2021}. This galaxy group consists of 20 galaxies distributed in a ${\sim}1 \, \rm deg^2$ region and are all within a structure ${\sim}$400 \kms wide. Its unusually high gas richness and non-Gaussian velocity dispersion distribution suggests a dynamically young group, still in its early stages of assembly. Mostly dominated by disk galaxies and few irregulars, it is an intermediate mass group, with dynamical mass of $\log_{10} (M_{\rm dyn}/\rm M_{\odot}) = 12.32$. We identified all galaxies within that overdensity, and found that all lie along the relation, suggesting that this group environment has not significantly affected the \hi\ content of these galaxies. We measured a slope of $0.515 \pm 0.019$ and an intercept of $-3.393^{+0.184}_{-0.185}$ for the structure, which is still consistent with the full sample. A more complete insight of the variation of the $D_{\rm HI}$\,--\,$M_{\rm HI}$ relation with large scale environments will be achieved with the full MIGHTEE-\hi\ survey area and redshift range.

\subsection{Evolution of the $D_{\rm HI}$\,--\,$M_{\rm HI}$ relation as a function of $z$}
Whereas previous studies were restricted to the very nearby Universe, our sample and the relation derived from it extends over a previously unexplored redshift range. To investigate the evolution of the $D_{\rm HI}$\,--\,$M_{\rm HI}$ as a function of redshift, we use our sample to test for any redshift dependence using similar approach as in \cite{ponomareva2021}. We divided our sample into two redshift bins with $z \leq 0.04$ and $z > 0.04$, and performed the linear fit as described in Section \ref{sec:the_relation} to each bin. The low-redshift subsample consists of 63 galaxies and spans four decades in \hi\ mass ranging from $7.4 \leq \log(M_{\rm HI}[\rm M_{\odot}]) \leq 10.4$, with a median mass of $1.65\times10^{9}\,\rm M_{\odot}$. The high-redshift subsample contains 141 galaxies covering 3 orders of magnitude in mass ($8.9 \leq \log(M_{\rm HI}[\rm M_{\odot}]) \leq 10.7$) and a median \hi\ mass of $5.65\times10^{9}\,\rm M_{\odot}$. Fig. \ref{fig:D_HI_M_HI:redshift} shows the best-fitting relation for each redshift bin. We observe marginal difference between the slope and intercept of the two subsamples, but the findings are consistent within the errors with the best-fitting relation of the full sample. The low-redshift bin has an intrinsic scatter of $0.052^{+0.006}_{-0.005}$ and observed scatter of 0.054 dex, which is consistent with \citet{Wang2016}, whose sample only reaches out to redshifts of $z \sim 0.03$. The high-redshift bin has a slightly larger scatter, both intrinsic ($\sigma_{\rm int} = 0.053^{+0.004}_{-0.003}$) and observed ($\sigma = 0.058$), but is consistent within the errors (see Table \ref{tab:bayesian_results}).

To investigate the effect of a possible mass bias, we performed the fit once again for each subsample, for a common \hi\ mass range of $ 8.9 \leq \log(M_{\rm HI}[\rm M_{\odot}]) \leq 10.4 $. We observe a similar trend in the values of intrinsic scatter, slope, and only marginal difference is present between the low and high-$z$ subsamples with the fits being consistent within the errors. 
This supports the conclusion that the relation features no obvious evolution with redshift, which is consistent with predictions from hydrodynamical cosmological simulations of galaxy formation and evolution, e.g. NEWHORIZON \citep{dubois2021}. However, this cannot be fully explored due to the still considerably small redshift range of our sample and will be further investigated with the full MIGHTEE-\hi{} survey.

\section{Summary and Conclusions}\label{sec:conclusion}

We have presented the \dhi\,--\,\mhi\ relation and measured the statistical properties of the homogeneous MIGHTEE-\hi\ Early Science data sample which contains 204 galaxies, spans 4 decades in \hi\ mass and extends to a redshift of $z \sim 0.084$. 

We measured galaxy \hi\ masses and used a novel 2D Gaussian fitting method to obtain the size of galaxy \hi\ discs. We have also classified the galaxies based on their optical morphology. The main results of our study are as follows:
\begin{itemize}[noitemsep,topsep=-8pt]
	\item[$\bullet$] For the first time, we are able to measure the intrinsic scatter of the $D_{\rm HI}$\,--\,$M_{\rm HI}$ relation and find that it is non-zero. Therefore, we conclude that the relation allows for an intrinsic variation of ${\sim} 10\%$  in $D_{\rm HI}$ at a given $M_{\rm HI}$. 
	\item[$\bullet$] All of the galaxies in our sample are found to lie on the $D_{\rm HI}$\,--\,$M_{\rm HI}$ relation, independent of morphological type. We also do not find any strong evidence for the environmental dependence when restricting our sample to the large group at $z \sim 0.044$.
	\item[$\bullet$] For the first time, we studied the $D_{\rm HI}$\,--\,$M_{\rm HI}$ relation beyond $z>0.03$. We find no evidence that the relation has evolved over the last one billion years similarly to the baryonic Tully-Fisher relation \citep{ponomareva2021}, suggesting that the galaxy discs have not undergone significant changes in their gas distribution and mean surface mass density over this period of time. This result is consistent with simulations of galaxy formation and evolution. For example, the latest results from the NEWHORIZON simulations show little-to-zero evolution over a Hubble time \citep{dubois2021}.
	
	\item[$\bullet$] The measured statistical properties of the relation (slope, observed scatter and zero point) are entirely consistent with the largest $z=0$ study by \cite{Wang2016}.\\
\end{itemize}
In conclusion, the successful study of the $D_{\rm HI}$\,--\,$M_{\rm HI}$ relation  using Early Science data from the MIGHTEE survey already substantiates MeerKAT's potential for transformational \hi\ science. The full MIGHTEE survey, covering 20 square degrees, will increase the explored volume out to $z\sim0.5$, and will be crucial to study the evolution of the $D_{\rm HI}$\,--\,$M_{\rm HI}$ relation as a function of redshift and large scale environments i.e. field vs groups, filaments and overdensities.

\section*{Acknowledgements}
We thank the anonymous referee for their quick and helpful
comments. Their careful reading of the manuscript significantly improved the quality of this paper.
The authors gratefully acknowledge Prof. Dr. J.M. van der Hulst and Francesco Sinigaglia for their useful comments and suggestions to improve the early drafts of this paper.

The MeerKAT telescope is operated by the South African Radio Astronomy Observatory, which is a facility of the National Research Foundation, an agency of the Department of Science and Innovation. We acknowledge the use of the ilifu cloud computing facility,\footnote{www.ilifu.ac.za} a partnership between the University of Cape Town, the University of the Western Cape, the University of Stellenbosch, Sol Plaatje University, the Cape Peninsula University of Technology and the South African Radio Astronomy Observatory. The ilifu facility is supported by contributions from the Inter-University Institute for Data Intensive Astronomy (IDIA – a partnership between the University of Cape Town, the University of Pretoria, the University of the Western Cape and the South African Radio astronomy Observatory), the Computational Biology division at UCT and the Data Intensive Research Initiative of South Africa (DIRISA). The authors acknowledge the Centre for High Performance Computing (CHPC), South Africa, for providing computational resources to this research project.

The Hyper Suprime-Cam (HSC) collaboration includes the astronomical communities of Japan and Taiwan, and Princeton University. The HSC instrumentation and software were developed by the National Astronomical Observatory of Japan (NAOJ), the Kavli Institute for the Physics and Mathematics of the Universe (Kavli IPMU), the University of Tokyo, the High Energy Accelerator Research Organization (KEK), the Academia Sinica Institute for Astronomy and Astrophysics in Taiwan (ASIAA), and Princeton University. Funding was contributed by the FIRST program from Japanese Cabinet Office, the Ministry of Education, Culture, Sports, Science and Technology (MEXT), the Japan Society for the Promotion of Science (JSPS), Japan Science and Technology Agency (JST), the Toray Science Foundation, NAOJ, Kavli IPMU, KEK, ASIAA, and Princeton University.

SHAR, RKK and SK are supported by the South African Research Chairs Initiative of the Department of Science and Technology and National Research Foundation. AAP acknowledges the support of the STFC consolidated grant ST/S000488/1. MJJ and AAP acknowledge support from the Oxford Hintze Centre for Astrophysical Surveys which is funded through generous support from the Hintze Family Charitable Foundation. IH, MJJ and AAP acknowledge support from the UK Science and Technology Facilities Council [ST/N000919/1]. BSF and MJJ would like to acknowledge support from the Africa-Oxford Visiting Fellows Programme. NM acknowledges the support of the LMU Faculty of Physics. EAKA is supported by the WISE research programme, which is financed by the Dutch Research Council (NWO). MG was partially supported by the Australian Government through the Australian Research Council's Discovery Projects funding scheme (DP210102103). IP acknowledges financial support from the Italian Ministry of Foreign Affairs and International Cooperation (MAECI Grant Number ZA18GR02) and the South African Department of Science and Technology's National Research Foundation (DST-NRF Grant Number 113121) as part of the ISARP RADIOSKY2020 Joint Research Scheme. IH acknowledges support from the South African Radio Astronomy Observatory which is a facility of the National Research Foundation (NRF), an agency of the Department of Science and Innovation. KS acknowledges support from the Natural Sciences and Engineering Research Council of Canada (NSERC).\\

This research has made use of NASA’s Astrophysics Data System Bibliographic Services. This research made use of Astropy,\footnote{http://www.astropy.org} a community-developed core Python package for Astronomy \citep{astropy:2013, astropy:2018}. 
\section*{Data Availability}
The MIGHTEE-\hi\ spectral cubes will be released as part of the first data release of the MIGHTEE survey, which will include maps of the sources discussed in this paper. The data release is described in Frank et al. in prep. Data products used in this work are available upon reasonable request to the corresponding author.

\bibliographystyle{mnras}
\bibstyle{mnras}
\bibliography{MIGHTEE_TFR}
\vspace*{1cm}

\bsp	
\label{lastpage}

\end{document}